\documentclass[12pt]{article}
\usepackage{amsfonts,bm}
\usepackage{graphicx}
\baselineskip
\offinterlineskip
\begin{document}

\begin{center}
{\bf Bose Operators, Coherent States, Truncation, Spin Coherent States, Lie Algebras 
and Spectrum}
\end{center}

\begin{center}
{\bf  Willi-Hans Steeb$^\dag$, Garreth Kemp$^\dag$, 
Yorick Hardy$^\ast$ and Dylan Durieux$^\dag$} \\[2ex]

$^\dag$
Department of Mathematics and Applied Mathematics, \\
University of Johannesburg, Auckland Park 2006, South Africa, \\
e-mail: {\tt steebwilli@gmail.com}\\
e-mail: {\tt garrethkemp@gmail.com}\\
e-mail: {\tt dylandurieux835@gmail.com}\\[2ex]

$^\ast$
School of Mathematics, University of the Witwatersrand, \\
Johannesburg, Private Bag 3, Wits 2050, South Africa, \\
e-mail: {\tt yorick.hardy@wits.ac.za}\\[2ex]
\end{center}

\strut\hfill

{\bf Abstract.} We study truncated Bose operators in 
finite dimensional Hilbert spaces. Spin coherent states 
for the truncated Bose operators and canonical coherent states
for Bose operators are compared.
The Lie algebra structure and the spectrum of the truncated Bose
operators are discussed. 
\strut\hfill

\section{Introduction}

Let $\hat b^\dagger$, $\hat b$ be Bose creation and annihilation
operators with the commutation relation $[b,b^\dagger]=I$, 
where $I$ is the identity operator and $\hat b|0\rangle=0|0\rangle$,
$\langle 0|0\rangle=1$ \cite{1}. Then for the operators
$$
\hat N=\hat b^\dagger \hat b, \quad \hat b^\dagger, \quad \hat b, \quad I
$$ 
we find the commutators 
$[\hat b^\dagger \hat b,\hat b^\dagger]=\hat b^\dagger$,
$[\hat b^\dagger \hat b,\hat b]=-\hat b$, $[\hat b^\dagger,\hat b]=-I$. 
All the other commutators are 0. It is well-known \cite{2} 
that a non-hermitian faithful representation by $3 \times 3$ matrices
is given by 
$$
\hat b^\dagger \hat b \to M_{22}=\pmatrix { 0 & 0 & 0 \cr 0 & 1 & 0 \cr 0 & 0 & 0 }, 
\quad
I \to M_{13}=\pmatrix { 0 & 0 & 1 \cr 0 & 0 & 0 \cr 0 & 0 & 0 }
$$
$$
\hat b^\dagger \to M_{23}=\pmatrix { 0 & 0 & 0 \cr 0 & 0 & 1 \cr 0 & 0 & 0 },
\quad
\hat b \to M_{12}=\pmatrix { 0 & 1 & 0 \cr 0 & 0 & 0 \cr 0 & 0 & 0 }
$$
since for the commutators we find
$$
[M_{22},M_{23}]=M_{23}, \quad
[M_{22},M_{12}]=-M_{12}, \quad
[M_{22},M_{13}]=0_3
$$
$$ 
[M_{23},M_{12}]=-M_{13}, \quad
[M_{23},M_{13}]=0_3, \quad
[M_{12},M_{13}]=0_3.
$$
Note that the matrices $M_{13}$, $M_{23}$, $M_{12}$ are nonnormal.
With 
$$
\pmatrix { \hat b_1^\dagger & \hat b_2^\dagger & \hat b_3^\dagger }M_{22}
\pmatrix { \hat b_1 \cr \hat b_2 \cr \hat b_3 } = 
\hat b_2^\dagger \hat b_2, \quad
\pmatrix { \hat b_1^\dagger & \hat b_2^\dagger & \hat b_3^\dagger }M_{13}
\pmatrix { \hat b_1 \cr \hat b_2 \cr \hat b_3 } = 
\hat b_1^\dagger \hat b_3
$$
etc we find the representation
$$
\hat b_2^\dagger \hat b_2, \quad \hat b_1^\dagger \hat b_3, \quad
\hat b_2^\dagger \hat b_3, \quad \hat b_1^\dagger \hat b_2
$$
of the Lie algebra.
\newline

Here we consider the four operators
$$
\hat N = \hat b^\dagger \hat b, \quad \hat b^\dagger+\hat b, \quad 
\hat b^\dagger-\hat b, \quad I
\eqno(1)
$$
and truncations into finite dimensional Hilbert spaces ${\mathbb C}^n$.
The four operators $\hat N$, $\hat b^\dagger+\hat b$, $\hat b^\dagger-\hat b$, $I$
form a basis of a Lie algebra. We obtain for the nonzero commutators
$$
[\hat b^\dagger b,b^\dagger+b]=\hat b^\dagger - \hat b, \quad
[\hat b^\dagger b,b^\dagger-b]=\hat b^\dagger + \hat b \quad
[\hat b^\dagger+b,b^\dagger-b]=2I.
$$
The Bargmann representation is
$$
\hat b^\dagger \leftrightarrow \zeta, \quad 
\hat b \leftrightarrow \frac{d}{d\zeta}, \quad
(\hat b^\dagger)^n|0\rangle \leftrightarrow \zeta^n, \quad 
|0\rangle \leftrightarrow 1
$$
with $n \in {\mathbb N}$ and the scalar product
$$
\langle f(\zeta)|g(\zeta)\rangle =
\frac1{\pi}\int_{\mathbb C} 
f(\zeta)\overline{g(\zeta)}e^{-|\zeta|^2} dx dy 
$$
where $\zeta=x+iy$.

Obviously the identity operator $I$ commutes 
with all other operators. Thus the Lie algebra generated by these operators is not 
semi-simple. The adjoint representation of this Lie algebra 
is given by
$$
\hat b^\dagger \hat b \mapsto 
\pmatrix { 0 & 0 & 0 & 0 \cr 0 & 0 & 1 & 0 \cr
0 & 1 & 0 & 0 \cr 0 & 0 & 0 & 0 }, \,\,\,
\hat b^\dagger + \hat b \mapsto 
\pmatrix { 0 & 0 & 0 & 0 \cr 0 & 0 & 0 & 0 \cr
-1 & 0 & 0 & 0 \cr 0 & 0 & 2 & 0 }, \,\,\,
\hat b^\dagger - \hat b \mapsto 
\pmatrix { 0 & 0 & 0 & 0 \cr -1 & 0 & 0 & 0 \cr
0 & 0 & 0 & 0 \cr 0 & -2 & 0 & 0 }
$$
with the identity operator mapping to the $4 \times 4$
zero matrix. Let $|n\rangle$, $|\beta\rangle$, $|\zeta\rangle$ 
be the number states $(n=0,1,\dots)$, canonical coherent states 
$(\beta \in {\mathbb C})$ and squeezed states 
$(\zeta \in {\mathbb C})$, respectively. Then we find 
for the operators given by (1) \cite{3}
$$
\langle n|\hat b^\dagger \hat b|n\rangle=n, \quad 
\langle \beta|\hat b^\dagger \hat b|\beta\rangle=\beta\beta^*, \quad
\langle \zeta|\hat b^\dagger \hat b|\zeta\rangle=\sinh^2(|\zeta|)
$$
$$
\langle n|(\hat b^\dagger + \hat b)|n\rangle=0, \quad 
\langle \beta|(\hat b^\dagger + \hat b)|\beta\rangle=2\Re(\beta), \quad
\langle \zeta|(\hat b^\dagger + \hat b)|\zeta\rangle=0
$$
$$
\langle n|(\hat b^\dagger-\hat b)|n\rangle=0, \quad 
\langle \beta|(\hat b^\dagger-\hat b)|\beta\rangle=-2\Im(\beta), \quad
\langle \zeta|(\hat b^\dagger-\hat b)|\zeta\rangle=0
$$
where $|\beta\rangle=D(\beta)|0\rangle$ and $|\zeta\rangle=S(\zeta)|0\rangle$
with the displacement operator $D(\beta)$ and squeezing operator $S(\zeta)$
given by
$$
D(\beta)=\exp(\beta \hat b^\dagger - \bar\beta \hat b), 
\qquad
S(\zeta)=
\exp\left(-\frac{\zeta}{2}(\hat b^\dagger)^2+\frac{\bar\zeta}{2}\hat b^2\right).
$$
Setting $\beta=e^{i\phi}\tan(\theta/2)$ with $\theta \in [0,\pi)$ and
$\phi \in [0,2\pi)$ we obtain
$$
\langle\beta|\hat b^\dagger \hat b|\beta\rangle=\tan^2(\theta/2), \quad
\langle\beta|(\hat b^\dagger+\hat b)|\beta\rangle= 
2\tan(\theta/2)\cos(\phi).
$$
If $\theta=0$, then $\langle \beta|\hat b^\dagger \hat b|\beta\rangle=0$
and $\langle \beta|(\hat b^\dagger+\hat b)|\beta\rangle=0$.
If $\theta=\pi/2$, then 
$\langle \beta|\hat b^\dagger \hat b|\beta\rangle=1$
and $\langle \beta|(\hat b^\dagger+\hat b)|\beta\rangle=2\cos(\phi)$.
These results can then be compared with the result from the truncated
Bose operators and spin coherent states.
\newline

We study the $n \times n$ matrices which arise in the
truncation of the four operators given by (1). 
Since the four operators given by (1) form a basis of
a Lie algebra we ask the question whether the $n \times n$
matrices from the truncation form a basis of a Lie algebra.
Furthermore we study the spectrum of the truncated operators.
Coherent states in a finite-dimensional Hilbert space 
have been studied by Miranowicz et al \cite{4,5}.
Utilizing the spin coherent states we find the expectation 
values for the truncated operators of $\hat b^\dagger \hat b$
and $\hat b^\dagger+\hat b$.
\newline

We mention that this set of operators given in (1) can also be 
considered for Fermi systems. 
Let $\hat c^\dagger$, $\hat c$ be Fermi creation and annihilation operators with 
$[\hat c,\hat c^\dagger]_+=I$, $[\hat c,\hat c]_+=0$, 
$[\hat c^\dagger,\hat c^\dagger]_+=0$, where $I$ is the identity operator 
and $\hat c|0\rangle=0|0\rangle$ $(\langle 0|0\rangle=1)$. 
Then the operators 
$$
\hat N = \hat c^\dagger \hat c, \quad \hat c^\dagger + \hat c, \quad 
\hat c^\dagger - \hat c, \quad I
$$
form a basis of a Lie algebra. We obtain the nonzero commutators
$$
[\hat c^\dagger \hat c,\hat c^\dagger+\hat c]=\hat c^\dagger - \hat c, \quad
[\hat c^\dagger \hat c,\hat c^\dagger - \hat c]=\hat c^\dagger + \hat c, \quad
[\hat c^\dagger+\hat c,\hat c^\dagger - \hat c]=2I-4\hat c^\dagger \hat c.
$$
Obviously the identity operator $I$ commutes with all other operators.
So the Lie algebra is not semi-simple. A representation of
these operators would be with $2 \times 2$ matrices
$$
\hat c^\dagger \hat c \mapsto \pmatrix { 0 & 0 \cr 0 & 1 }, \quad
\hat c^\dagger + \hat c \mapsto \pmatrix { 0 & 1 \cr 1 & 0 }, \quad
\hat c^\dagger - \hat c \mapsto \pmatrix { 0 & 1 \cr -1 & 0 }, \quad 
I \mapsto \pmatrix { 1 & 0 \cr 0 & 1 }.
$$

\section{Truncation and Lie Algebras}

To find the matrix representation of $\hat N=\hat b^\dagger \hat b$, 
$\hat b^\dagger+\hat b$,
$\hat b^\dagger-\hat b$ we are applying number states $|n\rangle$
$(n=0,1,\dots)$ with the properties 
$$
\hat b^\dagger|n\rangle=\sqrt{n+1}|n+1\rangle, \quad 
\hat b|n\rangle=\sqrt{n}|n-1\rangle.
$$
The number operator $\hat N$ is unbounded.
Since $\hat b^\dagger \hat b|n\rangle=n|n\rangle$
we obtain the infinite dimensional unbounded diagonal matrix
$\mbox{diag}(0,1,2,\dots)$.
Using the number states $|n\rangle$ we find the matrix representation of 
the unbounded operators $\hat B=\hat b^\dagger+\hat b$ as 
$$ 
\hat B=\hat b^\dagger + \hat b = 
\pmatrix { 0 & 1 & 0 & 0 & \dots \cr
1 & 0 & \sqrt2 & 0 & \dots \cr 0 & \sqrt2 & 0 & \sqrt3 & \dots \cr
0 & 0 & \sqrt3 & 0 & \dots \cr \vdots & \vdots & \vdots & 
\vdots & \ddots }.
$$
Finally $\hat b^\dagger-\hat b$ is given by the matrix
$$ 
\hat C=\hat b^\dagger - \hat b = 
\pmatrix { 0 & -1 & 0 & 0 & \dots \cr
1 & 0 & -\sqrt2 & 0 & \dots \cr 0 & \sqrt2 & 0 & -\sqrt3 & \dots \cr
0 & 0 & \sqrt3 & 0 & \dots \cr \vdots & \vdots & \vdots & \vdots
& \ddots }.
$$
The identity operator $I$ is represented by the infinite dimensional 
unit matrix. Now we truncate these infinite dimensional matrices. 
The truncation could also be found as follows.
Let $n \ge 1$ and $\{ |0\rangle\rangle,
|1\rangle, \dots, |n\rangle \}$ be an orthonormal 
basis in ${\mathbb C}^{n+1}$. Note that
$$
\sum_{\ell=0}^n |\ell\rangle \langle \ell| = I_{n+1}.
$$  
Consider the linear operators ($(n+1) \times (n+1)$ matrices)
$$
b_n=\sum_{j=1}^n \sqrt{j}|j-1\rangle \langle j|, \qquad
b_n^\dagger=\sum_{k=1}^n \sqrt{k}|k\rangle \langle k-1|
$$
with
$$
b_n^\dagger + b_n = \sum_{k=1}^n \sqrt{k}
(|k-1\rangle \langle k|+|k\rangle \langle k-1|.
$$
Then
\begin{eqnarray*}
b_n b_n^\dagger &=& \sum_{j=1}^n\sum_{k=1}^n 
\sqrt{j}\sqrt{k}|j-1\rangle\langle j|k\rangle 
\langle k-1| = \sum_{k=1}^n k|k-1\rangle\langle k-1| \\
b_n^\dagger b_n &=& \sum_{k=1}^n\sum_{j=1}^n
\sqrt{k}\sqrt{j}|k\rangle \langle k-1|j-1\rangle\langle j| 
=\sum_{j=1}^n j|j\rangle \langle j| 
\end{eqnarray*}
and we obtain the commutator 
$$
[b_n,b_n^\dagger]=b_n b_n^\dagger - b_n^\dagger b_n =
I_{n+1}-(n+1)|n\rangle \langle n|.
$$
If we select the standard basis as the orthonormal
basis we obtain the truncated matrices we consider in
the following.
\newline

Now we truncate the infinite-dimensional matrices to $n \times n$ 
matrices acting on the Hilbert space ${\mathbb C}^n$, where $n \ge 2$.
For $n=2$ we obtain the matrices 
$$
N_2=\pmatrix { 0 & 0 \cr 0 & 1 }, \quad 
B_2=\pmatrix { 0 & 1 \cr 1 & 0 }=\sigma_1, \quad 
C_2=\pmatrix { 0 & -1 \cr 1 & 0 }=-i\sigma_2
$$
when we truncate the infinite dimensional unbounded matrices 
$\hat b^\dagger \hat b$, $\hat b^\dagger+\hat b$ and $\hat b^\dagger-\hat b$, where
$\sigma_1$, $\sigma_2$, $\sigma_3$ denote the three Pauli spin matrices. 
For the commutator $[N_2,B_2]$, $[N_2,C_2]$, $[B_2,C_2]$ we find
\begin{eqnarray*}
[N_2,B_2] &=& \pmatrix { 0 & -1 \cr 1 & 0 }=C_2=-i\sigma_2 \cr
[N_2,C_2] &=& \pmatrix { 0 & 1 \cr 1 & 0 }=B_2=\sigma_1 \cr
[B_2,C_2] &=& 2\pmatrix { 1 & 0 \cr 0 & -1 }=2I_1 \oplus (-1)=2\sigma_3
\end{eqnarray*}
where $I_1$ is the $1 \times 1$ identity matrix and $\oplus$ denotes
the direct sum. For $n=3$ we obtain the $3 \times 3$ matrices 
$$
N_3=\pmatrix { 0 & 0 & 0 \cr 0 & 1 & 0 \cr 0 & 0 & 2 }, \quad 
B_3=\pmatrix { 0 & 1 & 0 \cr 1 & 0 & \sqrt2 \cr 0 & \sqrt2 & 0 }, \quad
C_3=\pmatrix { 0 & -1 & 0 \cr 1 & 0 & -\sqrt2 \cr 0 & \sqrt2 & 0 }
$$
when we truncate the infinite dimensional unbounded matrices 
$\hat b^\dagger \hat b$, $\hat b^\dagger+\hat b$, $\hat b^\dagger-\hat b$. 
We find the commutator $[N_3,B_3]$, $[N_3,C_3]$, $[B_3,C_3]$ as
\begin{eqnarray*}
[N_3,B_3] &=& \pmatrix { 0 & -1 & 0 \cr 1 & 0 & -\sqrt2 \cr 
0 & \sqrt2 & 0 } = C_3 \cr
[N_3,C_3] &=& \pmatrix { 0 & 1 & 0 \cr 1 & 0 & \sqrt2 \cr 0 & \sqrt2 & 0 } 
= B_3 \cr
[B_3,C_3] &=& 2\pmatrix { 1 & 0 & 0 \cr 0 & 1 & 0 \cr 0 & 0 & -2 } 
= 2I_2 \oplus 2(-2)
\end{eqnarray*}
where $I_2$ is the $2 \times 2$ identity matrix.
For $n=4$ we obtain the $4 \times 4$ matrices
$$
N_4=\pmatrix { 0 & 0 & 0 & 0 \cr 0 & 1 & 0 & 0 \cr
                 0 & 0 & 2 & 0 \cr 0 & 0 & 0 & 3 }, \quad
B_4=\pmatrix { 0 & 1 & 0 & 0 \cr 1 & 0 & \sqrt2 & 0 \cr
                 0 & \sqrt2 & 0 & \sqrt3 \cr 0 & 0 & \sqrt3 & 0 }, \quad
C_4=\pmatrix { 0 & -1 & 0 & 0 \cr 1 & 0 & -\sqrt2 & 0 \cr
                 0 & \sqrt2 & 0 & -\sqrt3 \cr 0 & 0 & \sqrt3 & 0 }
$$
when we truncate the infinite dimensional unbounded matrices $\hat b^\dagger \hat b$ and 
$\hat b^\dagger+\hat b$ to $4 \times 4$ matrices.
We find the commutators $[N_4,B_4]$, $[N_4,C_4]$, $[B_4,C_4]$ as 
\begin{eqnarray*}
[N_4,B_4] &=& 
\pmatrix { 0 & -1 & 0 & 0 \cr 1 & 0 & -\sqrt2 & 0 \cr
           0 & \sqrt2 & 0 & -\sqrt3 \cr 0 & 0 & \sqrt3 & 0 } = C_4 \cr
[N_4,C_4] &=& 
\pmatrix { 0 & 1 & 0 & 0 \cr 1 & 0 & \sqrt2 & 0 \cr
           0 & \sqrt2 & 0 & \sqrt3 \cr 0 & 0 & \sqrt3 & 0 } = B_4 \cr
[B_4,C_4] &=& 
2\pmatrix { 1 & 0 & 0 & 0 \cr 0 & 1 & 0 & 0 \cr
           0 & 0 & 1 & 0 \cr 0 & 0 & 0 & -3 } = 2I_3 \oplus 2(-3) 
\end{eqnarray*}
where $I_3$ is the $3 \times 3$ identity matrix. The commutators of a 
truncation for arbitrary $n$ is now obvious. We find 
$$
[N_n,B_n]=C_n, \quad 
[N_n,C_n]=B_n, \quad 
[B_n,C_n]=2I_{n-1} \oplus 2(-n+1)
$$
where $I_{n-1}$ is the $(n-1) \times (n-1)$ identity matrix.
Thus the commutation relations for 
$[\hat b^\dagger \hat b,\hat b^\dagger+\hat b]$, 
$[\hat b^\dagger \hat b,\hat b^\dagger-\hat b]$ are
preserved for the truncation to finite dimensional matrices, 
whereas the commutator $[\hat b^\dagger+\hat b,\hat b^\dagger-\hat b]$ is not
preserved, i.e. we do not find 2 times the $n \times n$
identity matrix $I_n$, but the direct sum of $2I_{n-1}$ and $2(-n+1)$.

\section{Truncation and Spin Coherent States}

Spin coherent states have been introduced by Radcliffe \cite{6} in 
1971. Arecchi et al \cite{7} also studied spin coherent states under
the name atomic coherent states in 1972.
The spin coherent states for spin-$\frac12$, spin-$1$, spin-$\frac32$
are given by
$$
|\theta,\phi\rangle_{1/2}=
\pmatrix { \cos(\theta/2) \cr \sin(\theta/2)e^{i\phi} }, \quad
|\theta,\phi\rangle_{1}=
\pmatrix { \cos^2(\theta/2) \cr \sqrt2 \cos(\theta/2)\sin(\theta/2)e^{i\phi} \cr
\sin^2(\theta/2)e^{2i\phi} },
$$
$$
|\theta,\phi\rangle_{3/2}=
\pmatrix { \cos^3(\theta/2) \cr \sqrt3\cos^2(\theta/2)\sin(\theta/2)e^{i\phi} \cr
\sqrt3 \cos(\theta/2)\sin^2(\theta/2) e^{2i\phi} \cr
\sin^3(\theta/2)e^{3i\phi} }.
$$
For spin-$\frac12$ we find
$$
\langle\theta,\phi|N_2|\theta,\phi\rangle=\sin^2(\theta/2), \quad
\langle\theta,\phi|B_2|\theta,\phi\rangle=\sin(\theta)\cos(\phi).
$$
As in the case for $\hat b^\dagger \hat b$ the expectation
value is independent of $\phi$ and as in the case for $\hat b^\dagger+\hat b$
the expectation value depends on $\phi$ in the form $\cos(\phi)$.
For $\theta=\pi/2$ we have 
$\langle\theta,\phi|B_2|\theta,\phi\rangle=\cos(\phi)$.
\newline

For spin-$1$ we obtain
$$
\langle\theta,\phi|N_3|\theta,\phi\rangle=
2\sin^2(\theta/2), \quad
\langle\theta,\phi|B_3|\theta,\phi\rangle=
\sin(\theta)(\sqrt2\cos^2(\theta/2)+2\sin^2(\theta/2))\cos(\phi).
$$
As in the case for $\hat b^\dagger \hat b$ the expectation
value is independent of $\phi$ and as in the case for $\hat b^\dagger+\hat b$
the expectation value depends on $\phi$ in the form $\cos(\phi)$.
For $\theta=\pi/2$ we obtain $\langle \theta,\phi|B_3|\theta,\phi\rangle=
\frac1{\sqrt2}(1+\sqrt2)\cos(\phi)$.
\newline

For spin-$\frac32$ we obtain
$$
\langle\theta,\phi|N_4|\theta,\phi\rangle=3\sin^2(\theta/2),
$$
$$
\langle\theta,\phi|B_4|\theta,\phi\rangle=
\sin(\theta)(\sqrt3 \cos^4(\theta/2)+3\sqrt2 \sin^2(\theta/2)\cos^2(\theta/2)
+3\sin^4(\theta/2))\cos(\phi).
$$
As in the case for $\hat b^\dagger \hat b$ the expectation
value is independent of $\phi$ and as in the case for $\hat b^\dagger+\hat b$
the expectation value depends on $\phi$ in the form $\cos(\phi)$.
For $\theta=\pi/2$ we obtain
$$
\langle \theta,\phi|B_4|\theta,\phi\rangle=
\frac{\sqrt3}{4}(1+\sqrt2\sqrt3+\sqrt3).
$$

\section{Truncation and Spectrum}

It is well known that the spectrum of the unbounded operator 
$\hat b^\dagger+\hat b$ is the whole real axis $\mathbb R$ \cite{8}.
Truncating the matrix representation of
the unbounded operator $\hat b^\dagger+\hat b$ up the $6 \times 6$ matrices
we obtain the symmetric matrices over $\mathbb R$
$$
B_2=\pmatrix { 0 & 1 \cr 1 & 0 }, \quad
B_3=\pmatrix { 0 & 1 & 0 \cr 1 & 0 & \sqrt2 \cr 0 & \sqrt2 & 0 }, \quad
B_4=\pmatrix { 0 & 1 & 0 & 0 \cr 1 & 0 & \sqrt2 & 0 \cr
0 & \sqrt2 & 0 & \sqrt3 \cr 0 & 0 & \sqrt3 & 0 }
$$
$$
B_5=\pmatrix { 0 & 1 & 0 & 0 & 0 \cr 1 & 0 & \sqrt2 & 0 & 0 \cr
           0 & \sqrt2 & 0 & \sqrt3 & 0 \cr 0 & 0 & \sqrt3 & 0 & \sqrt4 \cr
           0 & 0 & 0 & \sqrt4 & 0 }, \quad
B_6=\pmatrix { 0 & 1 & 0 & 0 & 0 & 0 \cr
               1 & 0 & \sqrt2 & 0 & 0 & 0 \cr
               0 & \sqrt2 & 0 & \sqrt3 & 0 & 0 \cr
               0 & 0 & \sqrt3 & 0 & \sqrt4 & 0 \cr
               0 & 0 & 0 & \sqrt4 & 0 & \sqrt5 \cr
               0 & 0 & 0 & 0 & \sqrt5 & 0 }.  
$$
We find the eigenvalues and eigenvectors of these matrices. 
Since the matrices are symmetric over
the real number the eigenvalues must be real.
Furthermore the sum of the eigenvalues must be 0
since the trace of the matrices is 0 and the eigenvalues
are symmetric around 0 \cite{9}. For $B_n$ with $n$ odd one of the 
eigenvalues is always 0. We order the eigenvalues from
largest to smallest. For $B_2$ we obtain 
the eigenvalues $1$, $-1$ with the eigenvectors
$$
\frac1{\sqrt2} \pmatrix { 1 \cr 1 }, \quad 
\frac1{\sqrt2} \pmatrix { 1 \cr -1 }.
$$
The eigenvalues of the matrix $B_3$ are
$\sqrt3$, $0$, $-\sqrt3$ with the corresponding 
unnormalized eigenvectors
$$
\pmatrix { 1 \cr \sqrt3 \cr \sqrt2 }, \quad
\pmatrix { 1 \cr 0 \cr -1/\sqrt2 }, \quad
\pmatrix { 1 \cr -\sqrt3 \cr \sqrt2 }.
$$
The eigenvalues of the matrix $B_4$ are
$$
\sqrt{3+\sqrt6}, \quad \sqrt{3-\sqrt6}, \quad 
-\sqrt{3-\sqrt6}, \quad -\sqrt{3+\sqrt6}
$$
with the corresponding unnormalized eigenvectors
$$
\pmatrix { 1 \cr \sqrt{3+\sqrt2\sqrt3 + 3} \cr 
           \sqrt2 + \sqrt3 \cr \sqrt{3+\sqrt2 \sqrt3} }, \quad 
\pmatrix { 1 \cr \sqrt{3-\sqrt2 \sqrt3} \cr 
           \sqrt2 - \sqrt3 \cr -\sqrt{3-sqrt2 \sqrt3} },
$$
$$ 
\pmatrix { 1 \cr -\sqrt{3-\sqrt2 \sqrt3 } \cr 
           \sqrt2 - \sqrt3 \cr \sqrt{3-\sqrt2 \sqrt3} }, \quad 
\pmatrix { 1 \cr -\sqrt{3+\sqrt2 \sqrt3 } \cr 
           \sqrt2 + \sqrt3 \cr -\sqrt{3+\sqrt2 \sqrt3} }.
$$
The eigenvalues of the matrix $B_5$ are 
$$
\sqrt{5+\sqrt{10}}, \quad \sqrt{5-\sqrt{10}}, \quad 
0, \quad 
-\sqrt{5-\sqrt{10}}, \quad -\sqrt{5+\sqrt{10}}
$$
with the corresponding unnormalized eigenvectors
The eigenvectors of $B_5$ are (for $\lambda=-\sqrt{5+\sqrt{10}},-\sqrt{5-\sqrt{10}},0,\sqrt{5-\sqrt{10}},\sqrt{5+\sqrt{10}}$)
$$
\pmatrix{1 \cr -\sqrt{5+\sqrt{10}} \cr (4+\sqrt{10})/\sqrt2\cr
\sqrt{5+\sqrt{10}}(2+\sqrt{10})/\sqrt6\cr
\sqrt{\frac23}(2+\sqrt{10})},\quad
\pmatrix{1 \cr -\sqrt{5-\sqrt{10}}\cr
(4-\sqrt{10})/\sqrt2\cr
\sqrt{5-\sqrt{10}}(2-\sqrt{10})/\sqrt6\cr
\sqrt{\frac23}(2-\sqrt{10})},\quad
\pmatrix{1 \cr 0 \cr -\sqrt{\frac12} \cr 0 \cr \sqrt{\frac38}},
$$
$$
\pmatrix{1\cr \sqrt{5-\sqrt{10}}\cr
(4-\sqrt{10})/\sqrt2\cr \sqrt{5-\sqrt{10}}(\sqrt{10}-2)/\sqrt6\cr
\sqrt{\frac23}(2-\sqrt{10})},\quad
\pmatrix{1\cr \sqrt{5+\sqrt{10}}\cr
(4+\sqrt{10})/\sqrt2 \cr \sqrt{5+\sqrt{10}}(2+\sqrt{10})/\sqrt6\cr
\sqrt{\frac23}(2+\sqrt{10})}.
$$
For $n \ge 6$ the eigenvalues have to be found numerically.
A numerical study of the case $n=6$ provides the
six eigenvalues 
$-0.61670659019259$, $0.61670659019259$, 
$-1.889175877753$, $1.889175877753$,
$-3.3242574335521$ and $3.3242574335521$.
Note that the eigenvalues are symmetric around 0.
A numerical study indicates that for large $n$
the largest eigenvalue grows like $\approx 2\sqrt{n}$.
Looking at the difference between the largest and 
the second largest eigenvalue a numerical study
indicates that for large $n$ one finds the scaling
law $2/n^{0.185}$.
\newline

\end{document}